\documentclass[prl,twocolumn,showpacs]{revtex4}
\usepackage{graphicx}

\begin{document}

\title{Optically induced energy and spin transfer in non-resonantly
coupled pairs of self-assembled CdTe/ZnTe quantum dots}

\author{T. Kazimierczuk}
 \email{Tomasz.Kazimierczuk@fuw.edu.pl}
\author{J. Suffczy\'nski}
\author{A. Golnik}
\author{J.A. Gaj}
\author{P. Kossacki}

\affiliation{%
Institute of Experimental Physics, University of Warsaw,\\
Ho\.za 69, 00-681 Warsaw, Poland }%

\author{P. Wojnar}
\affiliation{%
Institute of Physics, Polish Academy of Sciences,\\
Al. Lotnik\'ow 32/64, 02-688 Warsaw, Poland}%

\date{\today}

\begin{abstract}
Asymmetrical horizontal interdot coupling was demonstrated to exist in a
system of a~single plane of self-assembled CdTe/ZnTe quantum dots.
Photoluminescence excitation (PLE), second order photon correlation and
optical orientation were used as main experimental tools. Each individual
absorbing dot was identified by a sharp PLE resonance, assigned to neutral
exciton transition, while the corresponding emission contained several
excitonic transitions of different charge states in another single quantum
dot different from the absorbing one. Energy and spin transfer dynamics
were studied. A~high efficiency of spin transfer was found from optical
orientation in a~vertical magnetic field (70\%) as well as without the
magnetic field (40\%), in spite of a~significant anisotropic exchange
splitting of the absorbing dot. Coherent mechanism of linear-to-circular
polarization conversion was identified, with an efficiency (43\%) close to
the theoretical limit of 50\%.
\end{abstract}


\maketitle

One of the reasons of growing interest in semiconductor quantum dots is
related to perspectives of their application in quantum information \cite{loss98}.
The~potential of quantum dots (QDs) to generate single photons \cite{mich00} 
and entangled photon pairs \cite{akop06,stev06} have been demonstrated,
opening the way to use the QDs 
to produce optical qubits.
Polarization control of the emitted photons, a~prerequisite for quantum
information processing, has become 
an~important challenge, in view of anisotropic QD properties
\cite{sant02}. On the other hand, the electronic states 
of the QDs, especially their spin degrees of freedom, can be directly used
as~qubits \cite{vinc95}. In this respect, 
coherent manipulation of QD states \cite{stie01} and coupling between
qubits located in different QDs \cite{robl08} became an attractive field of
investigations. 
Both vertically \cite{tadd00} and horizontally \cite{rodt03} coupled QD
pairs have been studied. 
The research has been focused mainly on resonant coupling of identical
quantum dots, while studies of coupling between QDs with strongly different
ground state energy have been less frequent \cite{reis07}. 
In all cases the quantum dot pairs have been prepared by a~suitable growth
procedure aiming at positioning 
the corresponding dots at a~controlled distance from one another. 
In this work we demonstrate that even in a single plane of self assembled
quantum dots there are randomly created pairs of coupled dots. We
demonstrate the energy and spin transfer between such dots. Some
preliminary results of this study can be found in \cite{kazi07}.

The control of polarization of photons emitted by quantum dots may involve
such processes as circular-to-linear and linear-to-circular  polarization
conversion. Two different mechanisms may lead to such conversion, both
related to the QD anisotropy. The first one, population-based, takes place
in a~magnetic field oriented along the growth axis. Its occurence has been
confirmed experimentally, e.g., on individual InAs/GaAs QDs \cite{kowa08}. 
The~second conversion mechanism is based on coherent evolution of two
linearly polarized QD eigenstates between the excitation and the emission
events. It does not require a~magnetic field and does not work with light
linearly polarized along either of the QD 
principal axes. The optimal light polarization for this mechanism is at
$45^\circ$ with respect to the principal axes and the maximum possible
conversion efficiency is 50\%. This coherent mechanism of
circular-to-linear and linear-to-circular conversion has been observed
experimentally on CdSe/ZnSe QD ensembles \cite{asta06}. As the conversion
has been so far observed on QD ensembles only,
the resulting polarization degree is reduced due to averaging - the
reported values do not exceed a~few per cent. 
In this work we employed an efficient spin-conserving transfer between
coupled quantum dots in order to prepare and to probe specific well defined
quantum states of the quantum dot.

\begin{figure}
\includegraphics[width=8.5cm]{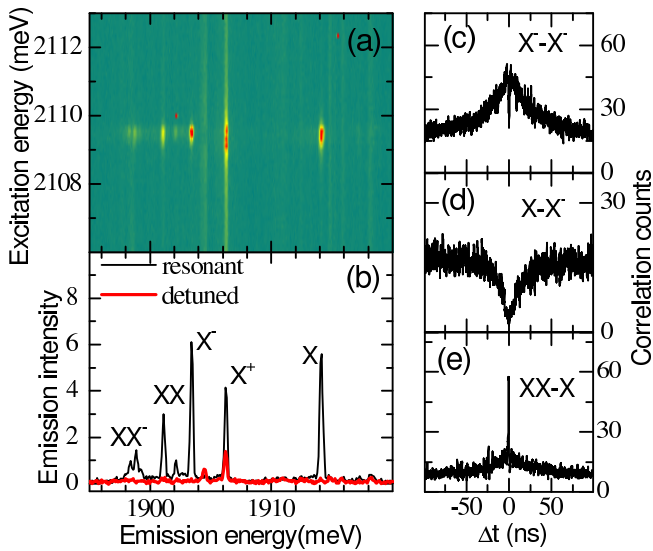}
\caption{\label{fig:ple_pl_korelacje}(color online) 
(a) PLE and (b) PL of a single QD measured under resonant excitation
(black line) and after detuning by 1~meV (red line). Correlation
histograms:
(c) X$^{-}$-X$^{-}$  autocorrelation, (d) anticorrelation between
different charge states (X-X$^{-}$) 
and~(e) XX-X  cascade.}
\end{figure}

The studied self-assembled CdTe/ZnTe 
QDs were grown by the tellurium desorption method \cite{tinj03}. The
samples were excited by a~CW tunable dye laser via 
a~microscope objective. Spectra of individual QDs were identified in the
low energy tail of 
the inhomogeneously broadened QD photoluminescence (PL) band. The
photoluminescence excitation (PLE) was scanned over the
maximum of the band. 
Weak above-barrier illumination was used in some of the experiments to
change the occupation of different
charge states. The most interesting features of PLE spectra of individual 
QDs are strong sharp (FWHM down to $90\ \mathrm{\mu eV}$)   resonances as
presented in Fig. \ref{fig:ple_pl_korelacje}(a).
Typical PL spectrum of a~single QD consists of a few distinct lines (Fig
1(b)).  
The pattern of the lines is unambiguously  
attributed to recombination of exciton (X), biexciton (XX) and charged
excitons (X$^{+}$, X$^{-}$). 
It was observed before in similar QD samples.
Several of investigated QDs exhibited an~additional transition identified
as originating from charged 
biexciton (XX$^{-}$). The energies of the absorbing resonances are
distributed in the range of maximum PL intensity of the QD emission band,
between 100 meV and 250 meV above the corresponding emission energies. No
correlation between absorption and emission energies was detected, as
reported in \cite{kazi07}. All the lines originating from each emitting QD
exhibit similar resonant behavior --- corresponding PLE
resonance energies are equal within accuracy of tens of $\mu$eV. These
facts allow us to formulate a~hypothesis that there are two different
coupled quantum dots involved: an absorbing- and an emitting one (it is
very unlikely that the resonance condition 
is fulfilled at the same energy for three different charge states).

The assignment of the PL lines was supported by measurements of QD
anisotropy. As expected, X and XX  emission lines exhibited a fine
structure splitting (FSS), forming doublets with two orthogonally linearly
polarized components. No FSS was detected for charged exciton emission. The
PLE resonances exhibited always the FSS doublet structure; therefore we
assigned them to neutral exciton transitions in the absorbing dots.
An unambiguous confirmation of the two coupled QDs model required
elimination of the possibility that the resonant absorption occurs in the
same (emitting) quantum dot in one of its charge states. 
If this had been the case, the observation of the two other ones would
require neutralization/charging of the QD during lifetime of the excited
state. To test this possibility, we performed photon correlation
measurements.
We used a Hanbury-Brown and Twiss setup described in detail in 
\cite{suff06}.  The sample was excited resonantly. As presented in Fig. 
\ref{fig:ple_pl_korelacje}(c)-(e), 
the measured histograms prove single- (or paired- in case of XX-X and
XX$^{-}$-X$^{-}$ cascades) photon emission. 
Widths of sharp features visible in autocorrelations (Fig.
\ref{fig:ple_pl_korelacje}c) 
and cross-correlations between cascade 
transitions (Fig. \ref{fig:ple_pl_korelacje}(e)) are about 300 ps and are
governed by the lifetime of the emitting states involved. 
In addition, we observed an~enhancement or a suppression of the signal in
the time range of tens of nanoseconds around zero delay.
The enhancement means that the two correlated transitions occur in the
same charge state, while for different charge states
a~suppression is observed. Moreover, the characteristic timescale of the
charge state variation in X$^{+}$-X$^{-}$  
correlation is about twice longer than that in X-X$^{-}$ or X-X$^{+}$. 
Thus the correlation measurements reveal an important difference between
two characteristic timescales of our system: that of emission decay
(hundreds of picoseconds) and that of the charge state variation (tens of
nanoseconds). This difference excludes a significant probability of the
charge state variation between the excitation and emission of a QD
excitonic state. We can therefore conclude at this point that the QDs are
not isolated but can interact 
with one another. Excitons created by resonant absorption in one QD relax
to emitting states in a neighboring one. 
Our analysis is supported by rate-equation model calculations, using 
capture rates of electron-hole pairs and of single carriers as fitting
parameters.
The obtained values are 0.2~ns$^{-1}$, 0.02~ns$^{-1}$ 0.03~ns$^{-1}$ for
pairs, electrons and holes respectively.

\begin{figure}
\includegraphics[width=9cm]{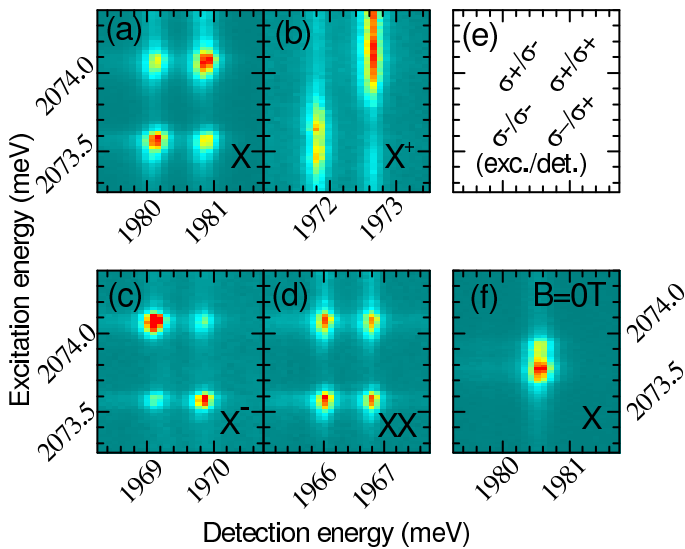}
\caption{\label{fig:mapy}(color online) PLE spectra at $B=4\ \mathrm{T}$
of transitions: a--X, b--X$^{+}$, 
c--X$^{-}$, d--XX. Each maximum corresponds to different combination of
excitation and detection circular polarizations
according to scheme e.  Map f presents PLE spectrum of X in $B=0\
\mathrm{T}$ for comparison. }
\end{figure}

Interdot spin transfer was studied first in magnetic field up to 6~T in
the Faraday configuration. In such a~field 
Zeeman effect (observed both for absorbing and emitting QD) dominates over
FSS and $S_z$ projection of the exciton spin is a~good quantum number. As a
result, four combinations 
of circular polarizations of excitation and detection were encoded as peak
positions in PLE maps
measured with no polarization resolution (Fig. \ref{fig:mapy}(e)). We
established that resonant PL intensity depends
on polarization of both excitation and detection as presented in Fig.
\ref{fig:mapy}(a)-(d).
Efficient spin transfer is visible for all excitonic transitions, except
XX, which is a~singlet state.
An interesting case of negative optical orientation occurs for X$^{-}$
transition. 
We interpret it in terms of electron-hole spin \emph{flip-flop} processes
in the emitting QD, following \cite{cort02}, where similar effect has been
observed before for negatively charged excitons in quantum dots pumped via
the wetting layer. 
The negative optical orientation would not be possible without relatively
fast spin relaxation of holes. In the experiments of \cite{cort02} the
relaxation ocurred in the wetting layer, while in our case the hole spin
may be depolarized during the energy relaxation process after the tunneling
of the photocreated exciton to the emitting dot.
The observed high degree of negative orientation (50\%)
proves that the \emph{flip-flop} process can be very efficient.

\begin{figure}
\includegraphics[width=9cm]{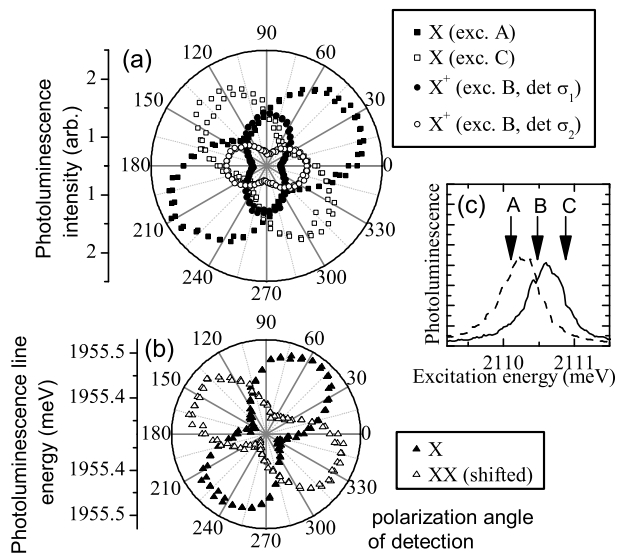}
\caption{\label{fig:anizotropie}Orientations of the anisotropy of the
absorbing (a) and emitting (b) dot in 
a~single QD pair. (a) Photoluminescence intensity at different fixed
detection energies detected with no polarization resolution (squares) or
circular polarization (circles), excited with rotating linear polarization.
(b) Averaged PL energy versus angle of rotating linear polarizer at
detection. (c) PLE spectra of the pair (solid and dashed lines --- linear
polarization along QD eigenaxes).
A,B,C -- excitation used for measurements presented in panel (a).}
\end{figure}

Spin transfer in the absence of magnetic field is more complex 
because of the FSS of neutral
exciton states. Figure \ref{fig:anizotropie} presents resulting anisotropy
effects in the absorbing and emitting QDs.
As expected, the splitting of the same value but opposite in sign is
visible for X and XX emission (Fig. \ref{fig:anizotropie}(b)).
 No splitting of charged exciton transitions was observed.
The orientation of measured QD anisotropy varied from dot to dot, which is
typical in II-VI compounds \cite{kude00}.

The relative anisotropy orientation of the two dots in the pair can be
determined from Fig. \ref{fig:anizotropie}(a,b). Measured
over many QD pairs, it exhibits a significant scatter but some correlation
is also present, probably due to local strain.
We did not observe transfer of linear polarization between dots.
On the other hand under certain conditions spin conservation leads to
circular polarization transfer. These facts favor tunneling mechanism of
the excitation transfer , as opposed to Foerster electromagnetic coupling.

\begin{figure}
\includegraphics[width=9cm]{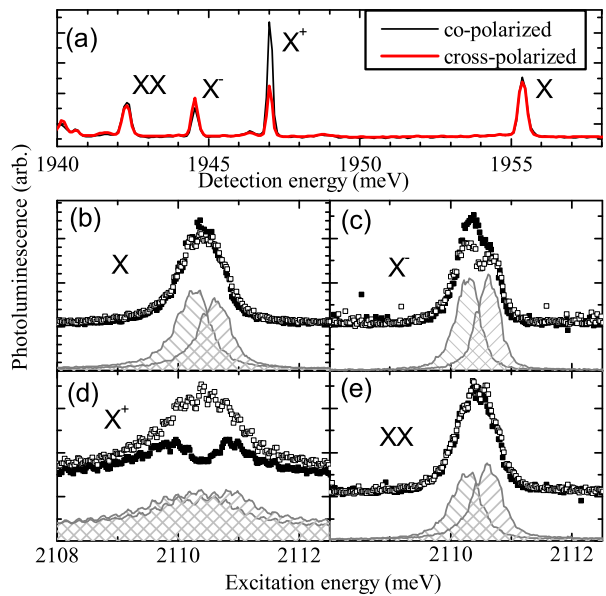}
\caption{\label{fig:konwersja}(color online) (a) Optical orientation of
a~QD with relatively broad resonance in circulary polarized excitation and
detection.
(b-e) PLE spectra of various transitions with excitation polarized
linearly at $45^\circ$ and different circular polarizations
of detection (points). Grey lines correspond to excitation light polarized
along principal axes of the absorbing QD (lines). The spectra were
vertically shifted for clarity.}
\end{figure}

For the study of spin conservation in absence of the field we selected
a~QD pair in which the resonance width was larger 
than the fine splitting in the absorbing QD. In such a case it is possible
to excite a~superposition of two
linear eigenstates, e.g.,
$\left(\left|x\right>+i\left|y\right>\right)/\sqrt{2}$. The
results are shown
in Figs. \ref{fig:anizotropie}(a) and \ref{fig:konwersja}. We found
a~significant circular polarization 
transfer to the two trion lines (Fig. \ref{fig:konwersja}(a)).

On the other hand, the two excited eigenstates of the absorbing QD
have slightly different energies due to in-plane anisotropy. Therefore the
relative phase $\varphi$ varies with time. 
As a consequence, the system undergoes 
oscillations between circular and $45^\circ$ linear polarization leading
to polarization conversion \cite{asta06}. 
We observed a~considerable
degree of circular polarization of emitted light after excitation with
linearly polarized light.
Again, the effect was observed only for trions as the anisotropy of the
emitting QD
enforces linear emission for X and XX lines.
 Maximum efficiency
was obtained if excitation was polarized at $45^\circ$ to the principal
axes of the absorbing QD as presented 
by circles in Fig. \ref{fig:anizotropie}(a). The effect was more
pronounced for X$^{+}$ than X$^{-}$ (Fig \ref{fig:konwersja}) 
due to a~larger overlap of the two linearly polarized absorption lines.
Difference in the sign of the conversion for the two trions
is consistent with previously discussed experiments of optical
orientation.

High efficiency of polarization conversion exceeding $40\%$ allowed us to
estimate transfer rate to be close to 
$\hbar / \delta_{\mathrm{FSS}}$, where $\delta_{\mathrm{FSS}}$ is FSS
value of the exciton in the absorbing dot. An efficiency of at least $40\%$
corresponds to the transfer rate between $0.8$ and 
$3.4\ \mathrm{ps}$.  This estimation remains in agreement with those based
on the width of the resonance, which yields  $0.7$, $3.1$, $2.2\
\mathrm{ps}$ for positive, neutral
and negative charge state respectively. The latter values may be
underestimated because of neglecting inhomogeneous broadening.  Similar
value of the tunneling time was observed in experiments on coupled 
CdSe/ZnSe quantum well and quantum dot system separated by $10\
\mathrm{nm}$ barrier \cite{jin03}. 
The distance between coupled QDs may be larger if the dots are
superimposed on 2D platelets
as reported in \cite{nguy07}.
We must stress here that the observed efficient polarization conversion is
possible due to  the comparable values of the FSS precession time and the
lifetime of the exciton in the absorbing dot. This lifetime is governed by
the interdot tunneling rate.

Concluding, we have shown evidence for inter-dot coupling in a single
plane of self-assembled CdTe/ZnTe quantum dots. The coupling was observed
as energy transfer between a neutral exciton state of the absorbing dot and
various excitonic states of the emitting dot. This transfer occurs via
tunneling of excitons between dots. High (at least 70\%) interdot spin
transfer efficiency has been demonstrated directly in vertical magnetic
field, while at zero field it was observed directly and evidenced through
high (above 40\%) efficiency of linear-to-circular polarization conversion.
Characteristic times of the system dynamics have been determined, ranging
from a few picoseconds (FSS precession and interdot tunneling) to tens of
nanoseconds (charge state variation).

\begin{acknowledgments}
This work was partially supported by the Polish Ministry of Science and
Higher Education as research grants in years 2006-2010 and by European
Project No. MTKD-CT-2005-029671.
\end{acknowledgments}


\begin{thebibliography}{20}
\expandafter\ifx\csname natexlab\endcsname\relax\def\natexlab#1{#1}\fi
\expandafter\ifx\csname bibnamefont\endcsname\relax
  \def\bibnamefont#1{#1}\fi
\expandafter\ifx\csname bibfnamefont\endcsname\relax
  \def\bibfnamefont#1{#1}\fi
\expandafter\ifx\csname citenamefont\endcsname\relax
  \def\citenamefont#1{#1}\fi
\expandafter\ifx\csname url\endcsname\relax
  \def\url#1{\texttt{#1}}\fi
\expandafter\ifx\csname urlprefix\endcsname\relax\def\urlprefix{URL }\fi
\providecommand{\bibinfo}[2]{#2}
\providecommand{\eprint}[2][]{\url{#2}}

\bibitem[{\citenamefont{Loss and DiVincenzo}(1998)}]{loss98}
\bibinfo{author}{\bibfnamefont{D.}~\bibnamefont{Loss}} \bibnamefont{and}
  \bibinfo{author}{\bibfnamefont{D.~P.} \bibnamefont{DiVincenzo}},
  \bibinfo{journal}{Phys.\ Rev.\ A} \textbf{\bibinfo{volume}{57}},
  \bibinfo{pages}{120} (\bibinfo{year}{1998}).

\bibitem[{\citenamefont{Michler et~al.}(2000)\citenamefont{Michler, Kiraz,
  Becher, Schoenfeld, Petroff, Zhang, Hu, and Imamoglu}}]{mich00}
\bibinfo{author}{\bibfnamefont{P.}~\bibnamefont{Michler}},
  \bibinfo{author}{\bibfnamefont{A.}~\bibnamefont{Kiraz}},
  \bibinfo{author}{\bibfnamefont{C.}~\bibnamefont{Becher}},
  \bibinfo{author}{\bibfnamefont{W.~V.} \bibnamefont{Schoenfeld}},
  \bibinfo{author}{\bibfnamefont{P.~M.} \bibnamefont{Petroff}},
  \bibinfo{author}{\bibfnamefont{L.}~\bibnamefont{Zhang}},
  \bibinfo{author}{\bibfnamefont{E.}~\bibnamefont{Hu}}, \bibnamefont{and}
  \bibinfo{author}{\bibfnamefont{A.}~\bibnamefont{Imamoglu}},
  \bibinfo{journal}{Science} \textbf{\bibinfo{volume}{290}},
  \bibinfo{pages}{2282} (\bibinfo{year}{2000}).

\bibitem[{\citenamefont{Akopian et~al.}(2006)\citenamefont{Akopian, Lindner,
  Poem, Berlatzky, Avron, Gershoni, Gerardot, and Petroff}}]{akop06}
\bibinfo{author}{\bibfnamefont{N.}~\bibnamefont{Akopian}},
  \bibinfo{author}{\bibfnamefont{N.~H.} \bibnamefont{Lindner}},
  \bibinfo{author}{\bibfnamefont{E.}~\bibnamefont{Poem}},
  \bibinfo{author}{\bibfnamefont{Y.}~\bibnamefont{Berlatzky}},
  \bibinfo{author}{\bibfnamefont{J.}~\bibnamefont{Avron}},
  \bibinfo{author}{\bibfnamefont{D.}~\bibnamefont{Gershoni}},
  \bibinfo{author}{\bibfnamefont{B.~D.} \bibnamefont{Gerardot}},
  \bibnamefont{and} \bibinfo{author}{\bibfnamefont{P.~M.}
  \bibnamefont{Petroff}}, \bibinfo{journal}{Phys.\ Rev.\ Lett}
  \textbf{\bibinfo{volume}{96}}, \bibinfo{pages}{130501}
  (\bibinfo{year}{2006}).

\bibitem[{\citenamefont{Stevenson et~al.}(2006)\citenamefont{Stevenson, Young,
  Atkinson, Cooper, Ritchie, and Shields}}]{stev06}
\bibinfo{author}{\bibfnamefont{R.~M.} \bibnamefont{Stevenson}},
  \bibinfo{author}{\bibfnamefont{R.~J.} \bibnamefont{Young}},
  \bibinfo{author}{\bibfnamefont{P.}~\bibnamefont{Atkinson}},
  \bibinfo{author}{\bibfnamefont{K.}~\bibnamefont{Cooper}},
  \bibinfo{author}{\bibfnamefont{D.~A.} \bibnamefont{Ritchie}},
  \bibnamefont{and} \bibinfo{author}{\bibfnamefont{A.~J.}
  \bibnamefont{Shields}}, \bibinfo{journal}{Nature}
  \textbf{\bibinfo{volume}{439}}, \bibinfo{pages}{179} (\bibinfo{year}{2006}).

\bibitem[{\citenamefont{Santori et~al.}(2002)\citenamefont{Santori, Fattal,
  Pelton, Solomon, and Yamamoto}}]{sant02}
\bibinfo{author}{\bibfnamefont{C.}~\bibnamefont{Santori}},
  \bibinfo{author}{\bibfnamefont{D.}~\bibnamefont{Fattal}},
  \bibinfo{author}{\bibfnamefont{M.}~\bibnamefont{Pelton}},
  \bibinfo{author}{\bibfnamefont{G.~S.} \bibnamefont{Solomon}},
  \bibnamefont{and} \bibinfo{author}{\bibfnamefont{Y.}~\bibnamefont{Yamamoto}},
  \bibinfo{journal}{Phys.\ Rev.\ B} \textbf{\bibinfo{volume}{66}},
  \bibinfo{pages}{045308} (\bibinfo{year}{2002}).

\bibitem[{\citenamefont{DiVincenzo}(1995)}]{vinc95}
\bibinfo{author}{\bibfnamefont{D.~P.} \bibnamefont{DiVincenzo}},
  \bibinfo{journal}{Science} \textbf{\bibinfo{volume}{270}},
  \bibinfo{pages}{255} (\bibinfo{year}{1995}).

\bibitem[{\citenamefont{Stievater et~al.}(2001)\citenamefont{Stievater, Li,
  Steel, Gammon, Katzer, Park, Piermarocchi, and Sham}}]{stie01}
\bibinfo{author}{\bibfnamefont{T.}~\bibnamefont{Stievater}},
  \bibinfo{author}{\bibfnamefont{X.}~\bibnamefont{Li}},
  \bibinfo{author}{\bibfnamefont{D.}~\bibnamefont{Steel}},
  \bibinfo{author}{\bibfnamefont{D.}~\bibnamefont{Gammon}},
  \bibinfo{author}{\bibfnamefont{D.}~\bibnamefont{Katzer}},
  \bibinfo{author}{\bibfnamefont{D.}~\bibnamefont{Park}},
  \bibinfo{author}{\bibfnamefont{C.}~\bibnamefont{Piermarocchi}},
  \bibnamefont{and} \bibinfo{author}{\bibfnamefont{L.}~\bibnamefont{Sham}},
  \bibinfo{journal}{Phys.\ Rev.\ Lett.} \textbf{\bibinfo{volume}{87}},
  \bibinfo{pages}{133603} (\bibinfo{year}{2001}).

\bibitem[{\citenamefont{Robledo et~al.}(2008)\citenamefont{Robledo, Elzerman,
  Jundt, Atature, Hogele, Falt, and Imamoglu}}]{robl08}
\bibinfo{author}{\bibfnamefont{L.}~\bibnamefont{Robledo}},
  \bibinfo{author}{\bibfnamefont{J.}~\bibnamefont{Elzerman}},
  \bibinfo{author}{\bibfnamefont{G.}~\bibnamefont{Jundt}},
  \bibinfo{author}{\bibfnamefont{M.}~\bibnamefont{Atature}},
  \bibinfo{author}{\bibfnamefont{A.}~\bibnamefont{Hogele}},
  \bibinfo{author}{\bibfnamefont{S.}~\bibnamefont{Falt}}, \bibnamefont{and}
  \bibinfo{author}{\bibfnamefont{A.}~\bibnamefont{Imamoglu}},
  \bibinfo{journal}{Science} \textbf{\bibinfo{volume}{320}},
  \bibinfo{pages}{772} (\bibinfo{year}{2008}).

\bibitem[{\citenamefont{Taddei et~al.}(2000)\citenamefont{Taddei, Colocci,
  Vinattieri, Bogani, Franchi, Fringeri, Lazzarini, and Salviati}}]{tadd00}
\bibinfo{author}{\bibfnamefont{S.}~\bibnamefont{Taddei}},
  \bibinfo{author}{\bibfnamefont{M.}~\bibnamefont{Colocci}},
  \bibinfo{author}{\bibfnamefont{A.}~\bibnamefont{Vinattieri}},
  \bibinfo{author}{\bibfnamefont{F.}~\bibnamefont{Bogani}},
  \bibinfo{author}{\bibfnamefont{S.}~\bibnamefont{Franchi}},
  \bibinfo{author}{\bibfnamefont{P.}~\bibnamefont{Fringeri}},
  \bibinfo{author}{\bibfnamefont{L.}~\bibnamefont{Lazzarini}},
  \bibnamefont{and} \bibinfo{author}{\bibfnamefont{G.}~\bibnamefont{Salviati}},
  \bibinfo{journal}{Phys.\ Stat.\ Sol.\ B} \textbf{\bibinfo{volume}{224}},
  \bibinfo{pages}{413} (\bibinfo{year}{2000}).

\bibitem[{\citenamefont{Rodt et~al.}(2003)\citenamefont{Rodt, Turck, Heitz,
  Guffarth, Engelhardt, Pohl, Strassburg, Dworzak, Hoffmann, and
  Bimberg}}]{rodt03}
\bibinfo{author}{\bibfnamefont{S.}~\bibnamefont{Rodt}},
  \bibinfo{author}{\bibfnamefont{V.}~\bibnamefont{Turck}},
  \bibinfo{author}{\bibfnamefont{R.}~\bibnamefont{Heitz}},
  \bibinfo{author}{\bibfnamefont{F.}~\bibnamefont{Guffarth}},
  \bibinfo{author}{\bibfnamefont{R.}~\bibnamefont{Engelhardt}},
  \bibinfo{author}{\bibfnamefont{U.~W.} \bibnamefont{Pohl}},
  \bibinfo{author}{\bibfnamefont{M.}~\bibnamefont{Strassburg}},
  \bibinfo{author}{\bibfnamefont{M.}~\bibnamefont{Dworzak}},
  \bibinfo{author}{\bibfnamefont{A.}~\bibnamefont{Hoffmann}}, \bibnamefont{and}
  \bibinfo{author}{\bibfnamefont{D.}~\bibnamefont{Bimberg}},
  \bibinfo{journal}{Phys.\ Rev.\ B} \textbf{\bibinfo{volume}{67}},
  \bibinfo{pages}{235327} (\bibinfo{year}{2003}).

\bibitem[{\citenamefont{Reischle et~al.}(2007)\citenamefont{Reischle, Beirne,
  Roßbach, Jetter, Schweizer, and Michler}}]{reis07}
\bibinfo{author}{\bibfnamefont{M.}~\bibnamefont{Reischle}},
  \bibinfo{author}{\bibfnamefont{G.~J.} \bibnamefont{Beirne}},
  \bibinfo{author}{\bibfnamefont{R.}~\bibnamefont{Roßbach}},
  \bibinfo{author}{\bibfnamefont{M.}~\bibnamefont{Jetter}},
  \bibinfo{author}{\bibfnamefont{H.}~\bibnamefont{Schweizer}},
  \bibnamefont{and} \bibinfo{author}{\bibfnamefont{P.}~\bibnamefont{Michler}},
  \bibinfo{journal}{Phys.\ Rev.\ B} \textbf{\bibinfo{volume}{76}},
  \bibinfo{pages}{085338} (\bibinfo{year}{2007}).

\bibitem[{\citenamefont{Kazimierczuk et~al.}(2007)\citenamefont{Kazimierczuk,
  Nowak, Suffczynski, Wojnar, Golnik, Gaj, and Kossacki}}]{kazi07}
\bibinfo{author}{\bibfnamefont{T.}~\bibnamefont{Kazimierczuk}},
  \bibinfo{author}{\bibfnamefont{S.}~\bibnamefont{Nowak}},
  \bibinfo{author}{\bibfnamefont{J.}~\bibnamefont{Suffczynski}},
  \bibinfo{author}{\bibfnamefont{P.}~\bibnamefont{Wojnar}},
  \bibinfo{author}{\bibfnamefont{A.}~\bibnamefont{Golnik}},
  \bibinfo{author}{\bibfnamefont{J.~A.} \bibnamefont{Gaj}}, \bibnamefont{and}
  \bibinfo{author}{\bibfnamefont{P.}~\bibnamefont{Kossacki}},
  \bibinfo{journal}{Acta\ Phys.\ Pol.\ A} \textbf{\bibinfo{volume}{112}},
  \bibinfo{pages}{321} (\bibinfo{year}{2007}).

\bibitem[{\citenamefont{Kowalik et~al.}(2008)\citenamefont{Kowalik, Krebs,
  Lemaitre, Gaj, and Voisin}}]{kowa08}
\bibinfo{author}{\bibfnamefont{K.}~\bibnamefont{Kowalik}},
  \bibinfo{author}{\bibfnamefont{O.}~\bibnamefont{Krebs}},
  \bibinfo{author}{\bibfnamefont{A.}~\bibnamefont{Lemaitre}},
  \bibinfo{author}{\bibfnamefont{J.~A.} \bibnamefont{Gaj}}, \bibnamefont{and}
  \bibinfo{author}{\bibfnamefont{P.}~\bibnamefont{Voisin}},
  \bibinfo{journal}{Phys.\ Rev.\ B} \textbf{\bibinfo{volume}{77}},
  \bibinfo{pages}{161305} (\bibinfo{year}{2008}).

\bibitem[{\citenamefont{Astakhov et~al.}(2006)\citenamefont{Astakhov,
  Kiessling, Platonov, Slobodsky, Mahapatra, Ossau, Schmidt, Brunner, and
  Molenkamp}}]{asta06}
\bibinfo{author}{\bibfnamefont{G.}~\bibnamefont{Astakhov}},
  \bibinfo{author}{\bibfnamefont{T.}~\bibnamefont{Kiessling}},
  \bibinfo{author}{\bibfnamefont{A.}~\bibnamefont{Platonov}},
  \bibinfo{author}{\bibfnamefont{T.}~\bibnamefont{Slobodsky}},
  \bibinfo{author}{\bibfnamefont{S.}~\bibnamefont{Mahapatra}},
  \bibinfo{author}{\bibfnamefont{W.}~\bibnamefont{Ossau}},
  \bibinfo{author}{\bibfnamefont{G.}~\bibnamefont{Schmidt}},
  \bibinfo{author}{\bibfnamefont{K.}~\bibnamefont{Brunner}}, \bibnamefont{and}
  \bibinfo{author}{\bibfnamefont{L.}~\bibnamefont{Molenkamp}},
  \bibinfo{journal}{Phys.\ Rev.\ Lett.} \textbf{\bibinfo{volume}{96}},
  \bibinfo{pages}{027402} (\bibinfo{year}{2006}).

\bibitem[{\citenamefont{Tinjod et~al.}(2003)\citenamefont{Tinjod, Gilles,
  Moehl, Kheng, and Mariette}}]{tinj03}
\bibinfo{author}{\bibfnamefont{F.}~\bibnamefont{Tinjod}},
  \bibinfo{author}{\bibfnamefont{B.}~\bibnamefont{Gilles}},
  \bibinfo{author}{\bibfnamefont{S.}~\bibnamefont{Moehl}},
  \bibinfo{author}{\bibfnamefont{K.}~\bibnamefont{Kheng}}, \bibnamefont{and}
  \bibinfo{author}{\bibfnamefont{H.}~\bibnamefont{Mariette}},
  \bibinfo{journal}{Appl.\ Phys.\ Lett.} \textbf{\bibinfo{volume}{82}},
  \bibinfo{pages}{4340} (\bibinfo{year}{2003}).

\bibitem[{\citenamefont{Suffczynski et~al.}(2006)\citenamefont{Suffczynski,
  Kazimierczuk, Goryca, Piechal, Trajnerowicz, Kowalik, Kossacki, Golnik,
  Korona, Nawrocki et~al.}}]{suff06}
\bibinfo{author}{\bibfnamefont{J.}~\bibnamefont{Suffczynski}},
  \bibinfo{author}{\bibfnamefont{T.}~\bibnamefont{Kazimierczuk}},
  \bibinfo{author}{\bibfnamefont{M.}~\bibnamefont{Goryca}},
  \bibinfo{author}{\bibfnamefont{B.}~\bibnamefont{Piechal}},
  \bibinfo{author}{\bibfnamefont{A.}~\bibnamefont{Trajnerowicz}},
  \bibinfo{author}{\bibfnamefont{K.}~\bibnamefont{Kowalik}},
  \bibinfo{author}{\bibfnamefont{P.}~\bibnamefont{Kossacki}},
  \bibinfo{author}{\bibfnamefont{A.}~\bibnamefont{Golnik}},
  \bibinfo{author}{\bibfnamefont{K.~P.} \bibnamefont{Korona}},
  \bibinfo{author}{\bibfnamefont{M.}~\bibnamefont{Nawrocki}},
  \bibnamefont{et~al.}, \bibinfo{journal}{Phys.\ Rev.\ B}
  \textbf{\bibinfo{volume}{74}}, \bibinfo{pages}{085319}
  (\bibinfo{year}{2006}).

\bibitem[{\citenamefont{Cortez et~al.}(2002)\citenamefont{Cortez, Krebs,
  Laurent, Senes, Marie, Voisin, Ferreira, Bastard, G\'erard, and
  Amand}}]{cort02}
\bibinfo{author}{\bibfnamefont{S.}~\bibnamefont{Cortez}},
  \bibinfo{author}{\bibfnamefont{O.}~\bibnamefont{Krebs}},
  \bibinfo{author}{\bibfnamefont{S.}~\bibnamefont{Laurent}},
  \bibinfo{author}{\bibfnamefont{M.}~\bibnamefont{Senes}},
  \bibinfo{author}{\bibfnamefont{X.}~\bibnamefont{Marie}},
  \bibinfo{author}{\bibfnamefont{P.}~\bibnamefont{Voisin}},
  \bibinfo{author}{\bibfnamefont{R.}~\bibnamefont{Ferreira}},
  \bibinfo{author}{\bibfnamefont{G.}~\bibnamefont{Bastard}},
  \bibinfo{author}{\bibfnamefont{J.-M.} \bibnamefont{G\'erard}},
  \bibnamefont{and} \bibinfo{author}{\bibfnamefont{T.}~\bibnamefont{Amand}},
  \bibinfo{journal}{Phys.\ Rev.\ Lett.} \textbf{\bibinfo{volume}{89}},
  \bibinfo{pages}{207401} (\bibinfo{year}{2002}).

\bibitem[{\citenamefont{Kudelski et~al.}(2000)\citenamefont{Kudelski, Golnik,
  Gaj, Mackowski, Karczewski, and Kossut}}]{kude00}
\bibinfo{author}{\bibfnamefont{A.}~\bibnamefont{Kudelski}},
  \bibinfo{author}{\bibfnamefont{A.}~\bibnamefont{Golnik}},
  \bibinfo{author}{\bibfnamefont{J.~A.} \bibnamefont{Gaj}},
  \bibinfo{author}{\bibfnamefont{S.}~\bibnamefont{Mackowski}},
  \bibinfo{author}{\bibfnamefont{G.}~\bibnamefont{Karczewski}},
  \bibnamefont{and} \bibinfo{author}{\bibfnamefont{J.}~\bibnamefont{Kossut}},
  in \emph{\bibinfo{booktitle}{Proceedings of the 25th International Conference
  on Physics of Semiconductors}} (\bibinfo{publisher}{Springer},
  \bibinfo{year}{2000}), p. \bibinfo{pages}{1249}.

\bibitem[{\citenamefont{Jin et~al.}(2003)\citenamefont{Jin, Zhang, Zheng, Kong,
  and Shen}}]{jin03}
\bibinfo{author}{\bibfnamefont{H.}~\bibnamefont{Jin}},
  \bibinfo{author}{\bibfnamefont{L.-G.} \bibnamefont{Zhang}},
  \bibinfo{author}{\bibfnamefont{Z.-H.} \bibnamefont{Zheng}},
  \bibinfo{author}{\bibfnamefont{X.-G.} \bibnamefont{Kong}}, \bibnamefont{and}
  \bibinfo{author}{\bibfnamefont{D.-Z.} \bibnamefont{Shen}},
  \bibinfo{journal}{Sol.\ Stat.\ Comm.} \textbf{\bibinfo{volume}{130}},
  \bibinfo{pages}{653} (\bibinfo{year}{2003}).

\bibitem[{\citenamefont{Nguyen et~al.}(2007)\citenamefont{Nguyen, Mackowski,
  Hoang, Jackson, Smith, and Karczewski}}]{nguy07}
\bibinfo{author}{\bibfnamefont{T.}~\bibnamefont{Nguyen}},
  \bibinfo{author}{\bibfnamefont{S.}~\bibnamefont{Mackowski}},
  \bibinfo{author}{\bibfnamefont{T.}~\bibnamefont{Hoang}},
  \bibinfo{author}{\bibfnamefont{H.}~\bibnamefont{Jackson}},
  \bibinfo{author}{\bibfnamefont{L.}~\bibnamefont{Smith}}, \bibnamefont{and}
  \bibinfo{author}{\bibfnamefont{G.}~\bibnamefont{Karczewski}},
  \bibinfo{journal}{Phys.\ Rev.\ B} \textbf{\bibinfo{volume}{76}},
  \bibinfo{pages}{245320} (\bibinfo{year}{2007}).

\end{thebibliography}
\end{document}